\documentclass{iaus}
\usepackage {graphicx}

\title[stellar feedback]{The role of stellar feedback in the formation
of galactic disks and bulges in a $\Lambda$CDM Universe.}

\author[Ceverino \& Klypin]{Daniel Ceverino and Anatoly Klypin}

\affiliation{Astronomy Department, New Mexico State University, Las
Cruces, NM, USA}

\pubyear{2007}
\volume{245}  
\pagerange{119--126}
\date{?? and in revised form ??}
\jname{Formation and Evolution of Galaxy Bulges}
\editors{A.C. Editor, B.D. Editor \& C.E. Editor, eds.}
\begin{document}

\maketitle

\begin{abstract}
Although supernova explosions and stellar winds happens at scales
bellow 100 pc, they affect the interstellar medium(ISM) and galaxy
formation. We use cosmological N-body+Hydrodynamics simulations of
galaxy formation, as well as simulations of the ISM to study the
effect of stellar feedback on galactic scales. Stellar feedback
maintains gas with temperatures above a million degrees. This gas
fills bubbles, super-bubbles and chimneys.  Our model of feedback, in
which 10\%-30\% of the feedback energy is coming from runaway stars,
reproduces this hot gas only if the resolution is better than 50
pc. This is 10 times better than the typical resolution in
cosmological simulations of galaxy formation. Only with this
resolution, the effect of stellar feedback in galaxy formation is
resolved without any assumption about sub-resolution physics. Stellar
feedback can regulate the formation of bulges and can shape the inner
parts of the rotation curve.  
\end{abstract}

Stellar feedback is the most difficult ingredient in galaxy formation
models. The basic idea is that massive stars inject energy, mass and
metals back to the ISM through stellar winds and supernova
explosions. It is crucial to understand where and how the energy from
massive stars is released back to the ISM. While a large fraction of
massive stars are found in stellar clusters, 10-30\% are found in the
field with high peculiar velocities (\cite[Gies 1987]{Gies}). The
current explanation of these runaway stars is that they were ejected
from stellar clusters (\cite{Blaauw}; \cite{Poveda}). Besides the
significant fraction of runaway stars, little attention has been paid
to their effect on galaxy formation.

We model runaway stars by adding a random velocity to a fraction of
the stellar particles, so \%10 of runaway stars have peculiar
velocities higher than 40 Km s$^{-1}$. 
The heating rate from massive stars is modeled as the rate of energy
injection from a single stellar population. This energy release
is nearly constant over 40 Myr (\cite{leitherer}).  
The cooling rates
must include cooling bellow $10^4$ degrees. The conditions of
molecular clouds should be included in order to reproduce the effect
of stellar feedback in the ISM. We first study the effect of stellar
feedback in the ISM, using simulations of a 4 Kpc piece of the ISM
with a 14 pc resolution (figure 1). Then, we check whether this picture
holds when the resolution is degraded to the resolution that our
cosmological simulations can achieve at high redshift.  We find that a
50-pc resolution plus the effect of runaway stars can reproduce a
3-phase ISM (Figure 2). Finally, we study the effect of stellar
feedback in galaxy formation at high redshift using cosmological
hydrodynamical simulations (Figure 3).

The simulations were performed using the Eulerian gasdynamics + N-body
Adaptive Refinement Tree (ART) code (\cite{Kravtsov}). This code
includes metallicity-dependent radiative cooling, UV heating, star
formation and stellar feedback.

\begin{figure}
\centerline{
\scalebox{1}{%
\includegraphics{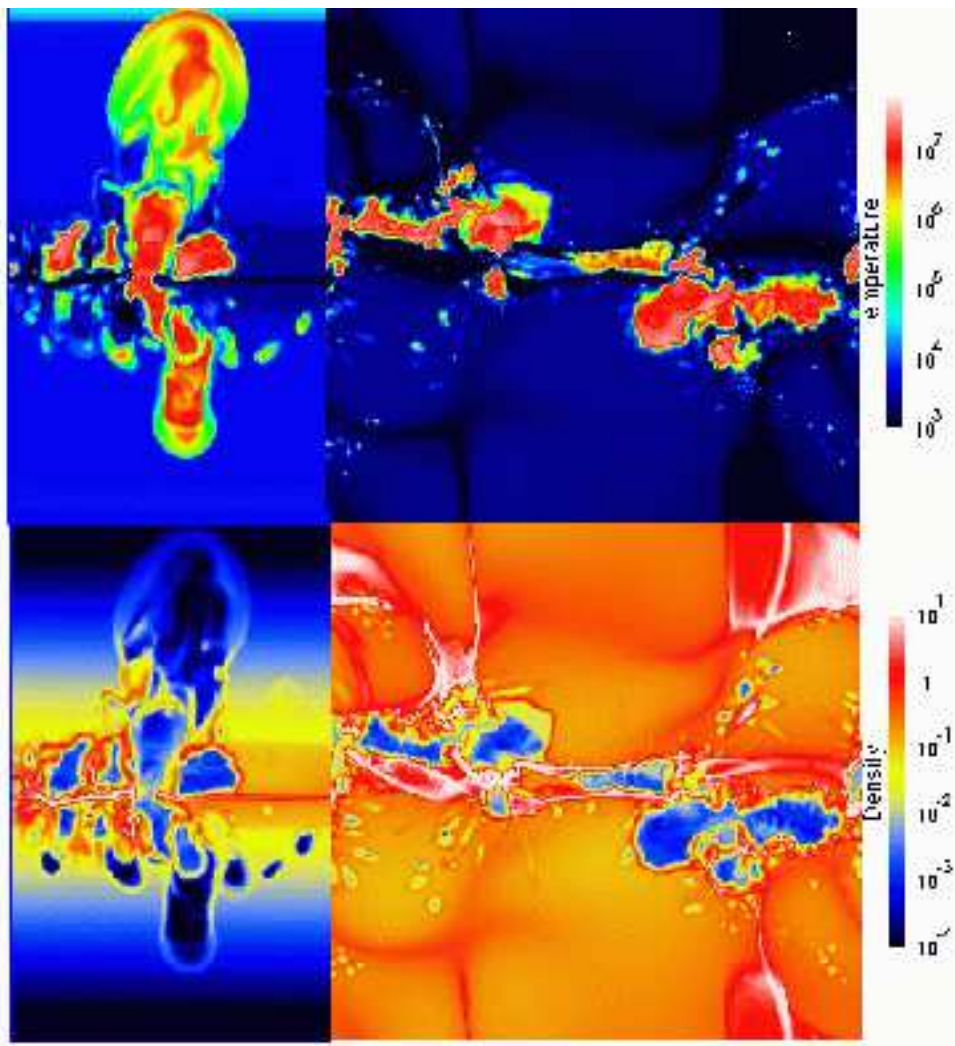}%
}
}
\caption{Density and temperature in slices
perpendicular to a galactic plane (left) and parallel (right). A
chimney filled with very hot and diffuse gas is created from the
energy input of stellar super-clusters.  Small bubbles in the field
are the result of the stellar feedback from runaway stars.
}
\end{figure}

\begin{figure}
\centerline{
\scalebox{0.6}{%
\includegraphics{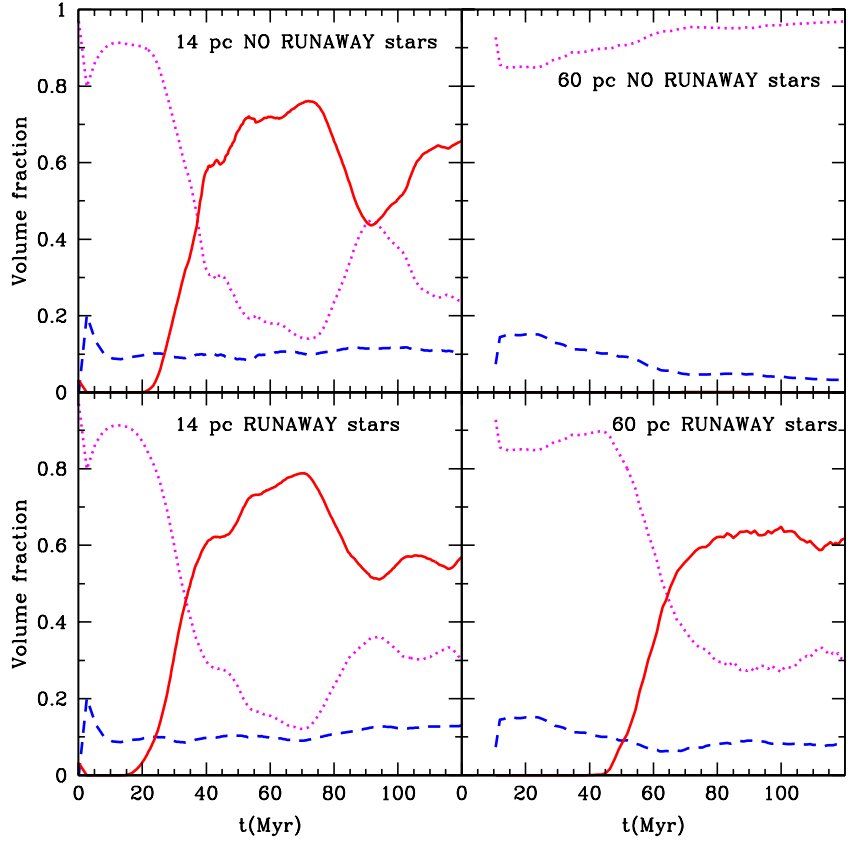}%
}
}
\caption{ Evolution of the volume occupied by different gas phases: $T
< 10^3 K$ (dashed line), $10^3 K < T < 10^4 K$ (dotted line), and $T >
10^4 K$ (full line). The simulations have a 1 Kpc box with four
different models. The hot phase does not develop in the low resolution
run without runaway stars (top right panel). If runaway stars are
included, the hot phase is recovered even at low resolution. As a
result, a fraction of runaway stars changes the thermodynamical
conditions of the ISM at low resolutions.}
\end{figure}

\begin{figure}
\centerline{
\scalebox{0.4}{%
\includegraphics{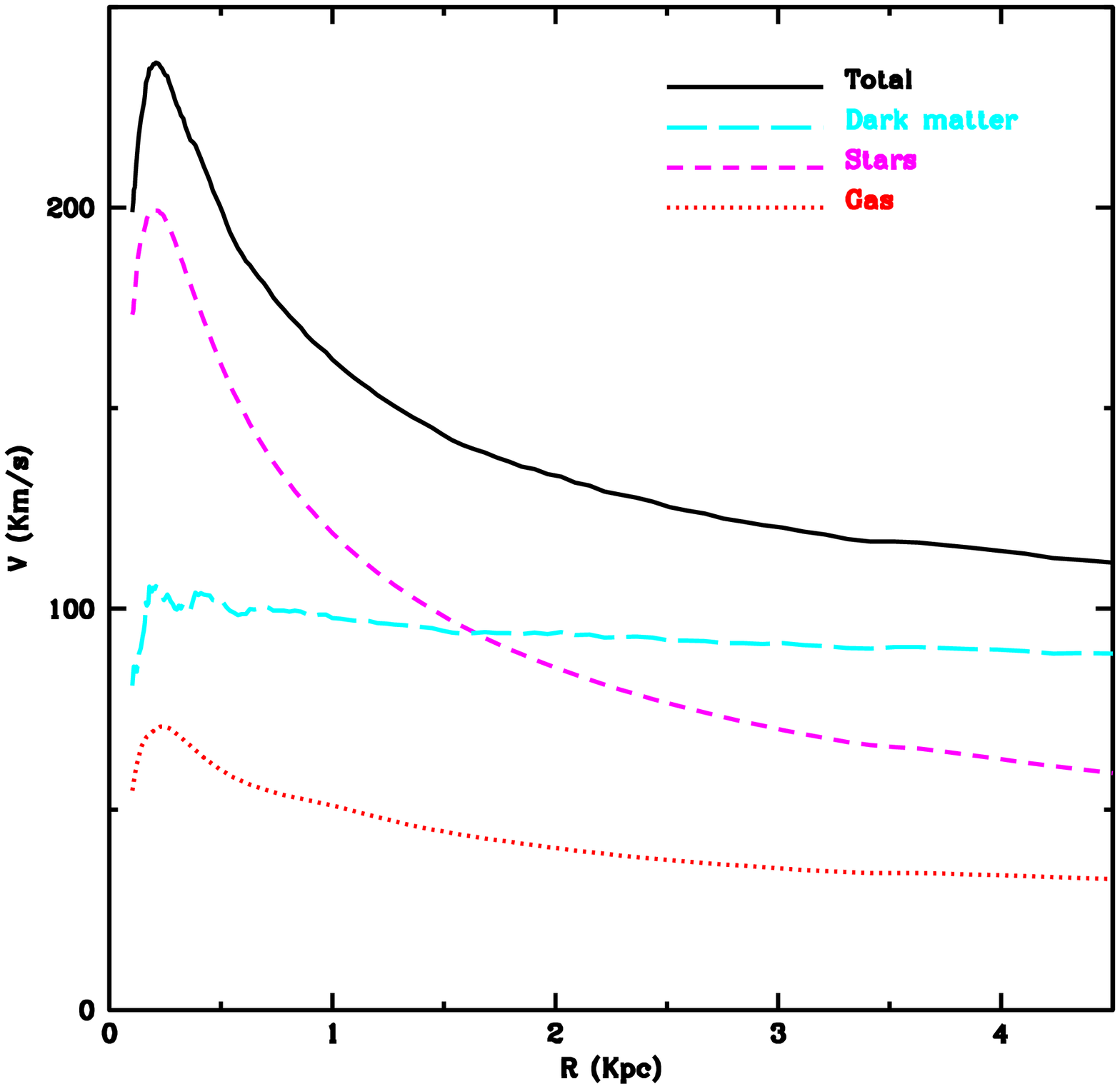}%
}
}

\centerline{
\scalebox{0.4}{%
\includegraphics{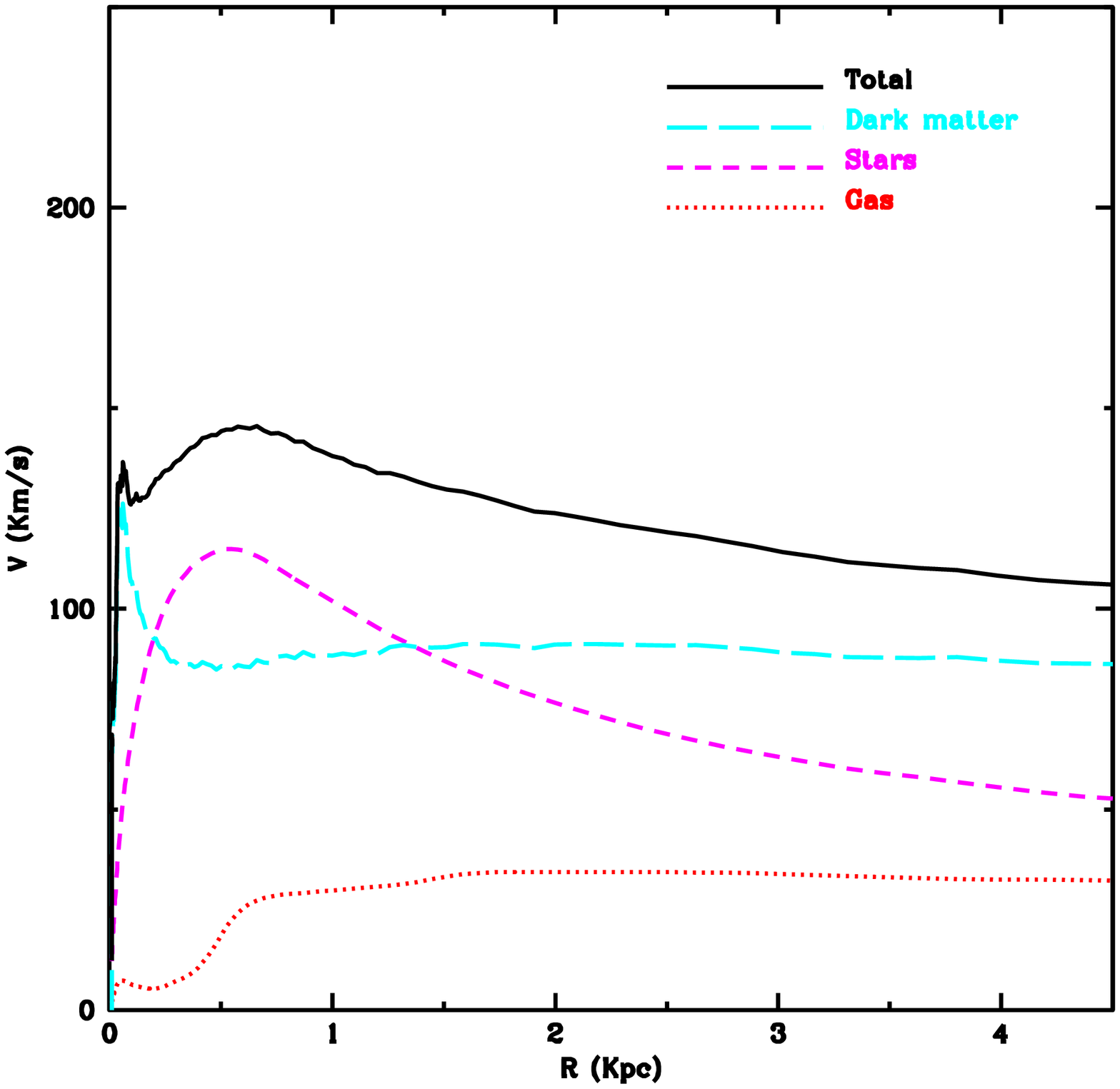}%
}
}

\caption{Circular velocity profile of a forming galaxy at redshift 6
for two different resolutions: 120 pc (top) and 60 pc (bottom). The
low-resolution case has the problem of a too massive bulge, which is
much weaker in the high-resolution case, where the stellar feedback is
more efficient and can regulate the growth of the bulge.}
\end{figure}


\clearpage

\begin{thebibliography}{}
\bibitem[Blaauw 1961]{Blaauw} Blaauw, A. 1961,
Bull. Astron. Inst. Neth. 15, 265
\bibitem[Gies 1987]{Gies}Gies, D. R. 1987, \textit{ApJS}, 64, 545
\bibitem[Kravtsov 2003]{Kravtsov}Kravtsov A., 2003, \textit{ApJ}
(Letters) 590, 1
\bibitem[Leitherer et al. 1999]{leitherer}Leitherer, C., et al.\ 1999,
\textit{ApJS} , 123, 3
\bibitem[Poveda et al. 1967]{Poveda}Poveda, A., Ruiz, J., \& Allen,
C. 1967, Bol. Obs. Tonantzintla Tacubaya, 4, 86
\end{thebibliography}
\end{document}